# Smart systems, the fourth industrial revolution and new challenges in distributed computing

Didier EL BAZ[1] and Li ZHU

*LAAS-CNRS, Université de Toulouse, CNRS Toulouse, France*

**Abstract.** Smart systems and the smart world concept are addressed in the framework of the fourth industrial revolution. New challenges in distributed autonomous robots and computing are considered. An illustration of a new kind of smart and reconfigurable distributed modular robot system is given. A prototype is also presented as well as the associated distributed algorithm.

## 1. Introduction

The first two industrial revolutions aimed essentially at increasing human productivity thanks to mechanization and use of steam engines (1760) or electric motors (1860). The third industrial revolution featured systematization and faster management thanks to automatic data treatment via computers (1960). Those three revolutions have changed profoundly the very nature of our societies, the way people work, or are staying, feed themselves, live, communicate as well as organize themselves. The fourth industrial revolution that aims at the fusion of physical, digital world and the Internet (2010) promises also stunning changes in the way people work and companies are organized. This revolution may lead to dramatic changes in the operation of companies and the factory of the future leading to more automatization, cooperation of robotics systems and workers, flexibility and better adaptation to client's demand. Workers and clerks will also probably be less mobile; many of them will work at home. High Tech companies are already hiring software developers around the world who work at home, several hundred miles from their offices, and use collaborative tools. The industrial revolution will have important consequences on transport due to the massive introduction of Artificial Intelligence (AI) on all transportation systems including intelligent highways in conjunction with probable reduction of travelers. This process will increase security in transport and diminish the production of greenhouse gases and pollution.

---

[1] Corresponding Author: Didier El Baz, LAAS-CNRS, 7 avenue du Colonel Roche, 31400, Toulouse, France; E-mail: elbaz@laas.fr IOS Press  Nieuwe Hemweg 6B, 1013 BG Amsterdam, The Netherlands; E-mail: bookproduction@iospress.nl.

Factory of the future will feature autonomous systems that can feed assembly lines and machines like 3D printers thanks to the progress in Artificial Intelligence (AI), as well as robots that can monitor activities or collaborate with workers for assembling complex machines like planes; Airbus has some projects like this. Factory of the future will feature also more efficient assembly lines like reconfigurable systems that can adapt to new goals or faulty situations in real time. This will lead to distributed, durable and economic robotic systems.

The fourth industrial revolution will feature also virtualization thanks to the progress in computer networks and development of cloud computing that will provide efficient storage and analytics services.

The concept of Internet of Things (IOT) [1], [2], [3], i.e., network of items embedded with sensors that are connected to the Internet has started to change the way people interact with the physical world just as the Internet has changed the way we communicate. Wireless sensor networks have a huge potential for applications ranging from manufacturing & agriculture to transportation, logistics, finance and security. We note also that the information about the physical environment is ubiquitously available: weather report, position of shuttles or trucks on roads, position of tractors in a farm, luggage at airport or people in the street. Information tends also to be embedded in the physical environment like in intelligent highways. This process has led also to the concept of Smart World (SW) and Smart Earth (SE), whereby the societies and physical world tend to be a large self-organizing system of growing complexity.

Let us note here that the fourth industrial revolution is more than a classical industrial revolution since it will have a direct impact on the cities and societies. One speaks commonly of smart cities [4] and smart world [5]. Amongst the many smart cities initiatives we can quote: Melbourne Australia, Rio de Janeiro Brazil, Madrid Spain, Amsterdam the Netherlands as well as important projects like Xiongan China and also Dijon France. Smart cities will feature in particular steering centers whereby all city data like water or energy demand as well as transportation data, public lighting, cleanliness, security and many more are gathered. Many of these projects involve the combined efforts of local administrations, universities and national or multinational companies and could have an important impact on citizen practice in the future leading to less waste of resources, less pollution and more security. Smart World has also a huge impact on health. IOT is now commonly used for monitoring health, screening, diagnose and monitoring of therapies. Many startups have appeared in this field related to health.

The design of new machines, new assembly lines and smart systems [6] in the framework of the factory of the future and the Smart World / Smart Earth leads to new challenges in computing. In particular, there are many algorithmic challenges in parallel computing related to efficiency and scalability. As an example, we note that the development of devices with massive parallelism or massive vectorization, e.g. computing accelerators like GPU and Xeon Phi, respectively permits one to encompass the relatively fast solution via parallel metaheuristics of difficult real world problems in modern manufacturing like complex flow shop scheduling problems [7]. Indeed, dynamic scheduling of new arriving jobs necessitates fast treatment. There are also many challenges in distributed computing related to efficiency, fault tolerance and security of cyber physical systems which are systems that are commonly seen as mechanisms controlled or monitored by computer-based algorithms, tightly integrated with the Internet and its users.

The problematic of intelligence is at the center of smart systems / smart world.

- One can consider systems with embedded intelligence that makes wide use of energy efficient computing accelerators like NVIDIA Jetson TX2 in order to monitor in real time a factory or analyze street contexts. Applications are for example in autonomous cars or autonomous conveyor in a factory;
- one can consider also systems with distributed intelligence like distributed autonomous robots that cooperate in order to implement a single task or a set of tasks and that can reconfigure (an example of smart conveyor will be presented in Section III).
- Finally, one can have hosted intelligence like intelligence in the cloud. The main advantage here is reduced cost and efficient analytics.

In this paper, we present some new challenges in computing in the context of reconfigurable autonomous robotics systems with distributed intelligence. This smart system is intended to be used for example for conveying, sorting or positioning micro parts in the in factory of the future.

Section 2 deals with related work in the domain of distributed autonomous robots. An example of reconfigurable modular robot for conveying or sorting micro parts is given in Section 3. The design of the distributed modular robot is presented in Section 3 and Section 4 presents distributed algorithms. Section 5 presents some conclusions and future work.

**2. Related work**

Robotic systems have taken an indispensable part in the human production process and many human industrial activities have been replaced by robots. In general, these robots are designed for specific-use according to mission requirements and environment constraints. Robotics systems are generally assigned to a specific mission that can hardly adapt to changes in the environment and tasks. It is also a huge investment to develop a new robotic system. Therefore, robotic systems that can change their structure based on modifications in their goals and evolution in the environment are welcome.

Some principles that had their origins in software engineering like reusability and reconfiguration are now used in robotics. With the joint development of electronics, MEMS and micro sensor technologies, devices like processors, sensors and actuators have become smaller and smaller. Control and communication systems can be integrated in one single module, which leads to the concept of modular robots. Modular Robots have gradually evolved into Modular Reconfigurable Robots (MRR). According to the reconfiguration process, MRR can also be divided into Modular Manual Reconfigurable Robots (MMRR) and Modular Self-Reconfigurable Robots (MSRR). MSRR can be traced back to the late 1980's when they were firstly introduced by Toshio Fukuda at the Science University of Tokyo with the name CEBOT [8] an abbreviation for 'cellular robotic system'. After more than 30 years of development, more than one hundred types of MSRR have been developed. Table 1 presents a list of some representative MSRR, their characteristics such as actuation, architecture, shape and degree of freedom (DOF) are displayed.

The concept of Catoms provide a good insight about what miniaturization of robots could give. Catoms belongs to the Claytronics project which was conducted by Seth C. Goldstein [13]. The novel feature of Catoms is their ability to reconfigure via electronic magnets that are arranged around the robots, by controlling the states of electro-magnets, a module can move and connect to other modules.

Blinky blocks system was also designed by Seth Copen Goldstein's team at Carnegie Mellon University [19], it is a modular distributed execution environment composed of centimeter-size blocks that are attached to each other using magnets. It is mainly used to test the distributed programming and communications.

M-blocks system was developed by John W. Romanishin, Kyle Gilpin, Daniela Rus at MIT [23], this system presents a novel self-assembling, self-reconfiguring cubic robot that uses pivoting to change its geometry.

Meanwhile, distributed algorithms for MSRR have also been designed. In order to facilitate the task of programming ATRON, U. P. Schultz [26] presented a concept of distributed control diffusion: distributed queries are used to identify modules that play a specific role in the robot, and behaviors that implement specific control strategies are diffused throughout the robot based on these role assignments. Kamimura et.al. developed both centralized and decentralized control method for M-TRANN III [27]. W M Shen et.al. have presented a biologically inspired approach to distributed collaboration between the physically coupled modules for CONRO. This approach was used to accomplish global effects such as locomotion and reconfiguration [28]. Miao et al. [29] proposed a distributed algorithm for enveloping an object inside a hexagonal lattice environment based on local communications among neighboring modules and between modules and the lattice node containing a target object.

**Table 1.** Some representative MSRR

| Year | Modular robot | Actuation | Architecture | Shape | DOF | Ref. |
|---|---|---|---|---|---|---|
| 1988 | CEBOT I | SMA | Chain | Cuboid | 3D | [8] |
| 1999 | CONRO | SMA | Chain | Cuboid | 3D | [9] |
| 2000 | Crystalline | DC motor | Lattice | Cuboid | 2D | [10] |
| 2002 | PolyBot II, III | SMA | Chain | Cubical | 3D | [11] |
| 2004 | ATRON | DC motor | Lattice | Octagonal | 3D | [12] |
| 2005 | Catom | Voltage | Free form | Cylindrical | 2D | [13] |
| 2005 | Molecube | Current | Hybrid | Cubical | 3D | [14] |
| 2005 | Programmable parts | DC motor | Lattice | Triangular | 2D | [15] |
| 2008 | M-TRAN III | DC motor | Hybrid | Semicylnd | 3D | [16] |
| 2010 | Sambot | Servo motor | Lattice | Cubical | 3D | [17] |
| 2010 | Pebbles | Random move | Lattice | Cubical | 2D | [18] |
| 2011 | Blinky blocks | Manual | Lattice | Cubical | 3D | [19] |
| 2012 | Smart Blocks | EP magnet | Lattice | Cubical | 3D | [20] |
| 2012 | SMORES | DC motor | Hybrid | Cubical | 3D | [21] |
| 2013 | USS | Servo motor | Chain | Cubical | 3D | [22] |
| 2015 | M-blocks 3D | EP magnet | Lattice | Cubical | 3D | [23] |
| 2016 | Larva-Bot | Servo motor | Lattice | Cubical | 3D | [24] |
| 2016 | Pitch-pitch | Servo motor | Chain | Cubical | 3D | [25] |

## 3. An example of distributed autonomous robot

This section briefly presents our contribution to the domain of distributed autonomous robots in the framework of the fourth industrial revolution.

### 3.1. Design principles

In order to illustrate challenges in distributed autonomous robots and distributed computing in the framework of the fourth industrial revolution, we introduce the DILI modular system of the Smart Surface [6], [30] and Smart Blocks projects [20].

### 3.2. Structure of DILI

As shown in Figure 1, the DILI module has a cubic shape. Each module has four work surfaces devoted to both motion and docking. Each module has six big holes for Semi Electro Permanent Magnets (SEP) and four small holes for NdFeBs, they are placed in two layers.

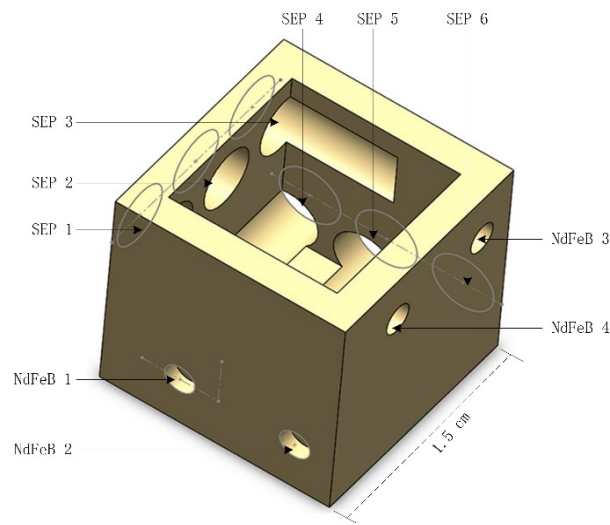

**Figure 1.** Global design of DILI module.

When moving, the surface with NdFeB faces the surface with SEP at the same height, as shown in Figure 2. The two SEPs and three NdFeBs form a linear motor. Each module can move by itself along other modules or can be driven by other modules. Placing the SEPs and NdFeBs in two layers guarantees that every sides has the ability of motion and permits one to make full use of the space and reduce the volume. The basic DILI module is built via 3D printer with Poly Lactic Acid (PLA).

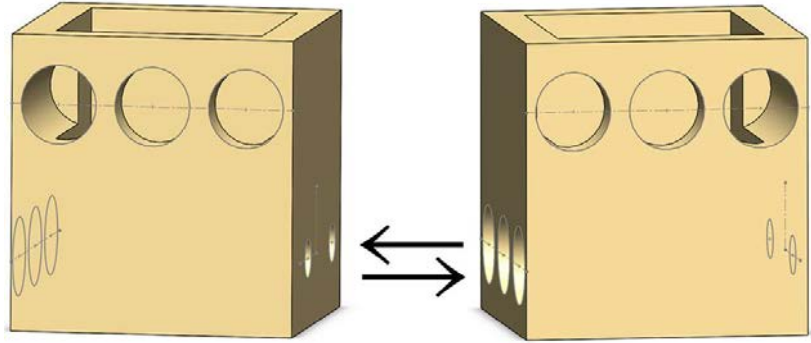

**Figure 2.** DILI modules: NdFeB faces SEP.

Six steps are needed to achieve a whole module motion along a distance of one module. At each step, only one SEP changes its polarity and the system only needs energy when module moves, after that, it does not need any energy to fasten to another module, since modules fasten to each other via the permanent magnetic field generated by Alnico5 and NdFeB. The linear motor can be very tiny; another advantage is that it can move in any direction, thanks to the controllability of the polarity of Alnico5. The speed of the DILI module is around 12 mm/s on several materials like cement, glass, paper and wood. We note that the speed is almost indistinguishable on the different surfaces. The reason that the surface has slight effects on the speed is that the movement of DILI is controlled by pulses. The movement is stepping forward rather than linear. When the module moves from one position to the next position, the initial speed is zero, the final speed is also zero, there is no speed accumulation. So, the previous move did not affect the next move. Figure 3 displays a motion test. A video of our test is presented in [31]; in particular, we observe a particularly smooth motion of the module.

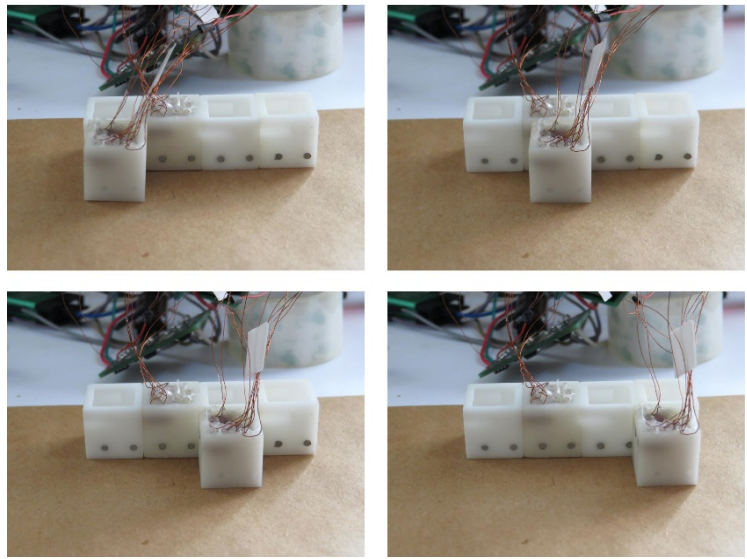

**Figure 3.** DILI module motion test.

DILI's new concept of distributed modular robot system has several advantages.
- It only needs one system to achieve both motion and docking, this permits one to reduce the complexity of the module, as well as its volume and to save energy.
- It does not require continuous energy supply for motion or fastening.
- The module can be made at a small scale. No space is wasted.
- If one of the modules is broken, then it can be driven by other modules.

**4. Challenges in computing**

In the recent years parallel and distributed computing have started to converge thanks to advances in high bandwidth networks and devices with massive parallelism like GPU and Intel Xeon Phi. In particular, the concept of parallelism has become essential; we have seen the development of multicore CPUs. Devices like GPUs, and MIC as well as heterogeneous computing prevail. With the advent of massive parallelism, fault tolerance has become a very important topic. Today, supercomputers typically have hundred thousands or millions computing cores. New concepts have also emerged like cloud computing, volunteer computing, and Peer-to-Peer (P2P) as well as the broader concept of Internet computing that covers all computing paradigms over the Internet.

IOT and Smart World present many applications where HPC is needed like computing vehicle tours in logistics and dynamic scheduling in factory of the future [2]. In particular, we need new combined requirements for integrated IOT and HPC or Smart World and HPC [32]. We need also fault-tolerant scalable parallel and distributed algorithms like reconfigurable distributed algorithms and parallel deep learning as well as new concepts for Big data treatment.

The introduction of sensor networks that collect usage or elements of context will permit to deal with the inherent complexity of the physical world that leads to many HPC problems. In particular, sensor processing, environment modeling, actuator networks with distributed intelligence will facilitate treatment.

IoT applications and Smart Earth can also take benefit of devices like GPU, MIC, FPGA, and concepts like cloud and volunteer computing that revisit distributed computing. Also, new parallel / distributed programming paradigms are needed to facilitate programming like in the programmable matter and blinky blocks contexts [13], [19].

Designing algorithms for the DILI distributed autonomous robot was an opportunity to address several issues like application deployment, cooperation, communication, reconfiguration, self-organization, asynchronous behavior versus synchronicity and scalability. The current distributed algorithm that is implemented is an improved version of the algorithm in [20], it is based on a heuristics method for solving combined optimization problems. It mainly tries to minimize the number of block motions in order to quickly set up a set of modules with shortest path between two points on a surface typically, the input and the output of the smart conveyor.

Without loss of generality we assume that the initial set of blocks is connected and located around the input. Our distributed algorithm has three following steps:
- Perform a distributed asynchronous election of a module that could be a good candidate for possible motion towards the output.

- Define a 3x3 square domain centered at the selected module and consider all the possible motion towards the output of the nine modules in this domain.
- when all the modules have been considered, then go to step 1.

This distributed algorithm can be used in a quasi-static way to reconfigure the modular system according to new goals, e.g. new output of the conveyor. We have also designed and developed a Python Simulator of Smart Modules (SSM). In particular, the SSM simulator has permitted us to test the distributed algorithm.

**5. Conclusions and future work**

IOT, smart world and smart earth domains are rich in applications needing High Performance Computing (HPC) like modern logistic applications where we need to solve rapidly difficult combinatorial optimization problems. We need for example to compute very fast new routes for trucks or cars like in the traveling salesman problems. This permits the companies to get big savings in gas and to reduce pollution. In the domain of autonomous transports, efficient parallel deep learning algorithms are also needed to analyze context in real time and forecast threats. Here people may benefit a lot of Graphics Processing Units (GPUs). In the dynamic context of the factory of the future where new jobs can arrive fast, we need also HPC. We need for example to solve very fast or in real time energy efficient dynamic flexible flow shop scheduling problems.

Smart distributed systems like smart autonomous robots present also many situations where scalable and fault tolerant distributed algorithms are needed. For this kind of application, cloud computing appears as an alternative for controlling robots and a way to perform statistics on items manipulated by the smart system and faults in the distributed system like in reconfigurable distributed smart conveyors. New paradigms of programming are needed here in order to facilitate programming like in programmable matter and blinky blocks contexts.

On what concerns the DILI modules, future directions of research will concern the design of secure, fault tolerant and scalable distributed algorithms for reconfiguration as well as nano communications protocols.

**References**

[1] H. Ning, *Unit and Ubiquitous Internet of Things*, CRC Press, ISBN 9781466561663, April 2013
[2] D. El Baz, IoT and the Need for High Performance Computing, Invited lecture, in *Proceedings of the International Conference on Identification, Information and Knowledge in The Internet of Things* (IIKI2014), IEEE CPS, 17-18 October 2014, Beijing China; 1-6,
[3] K. Hwang, G. Fox, J. Dongarra, *Distributed and Cloud Computing: From Parallel Processing to the Internet of Things*, Morgan Kaufmann Publishers Inc., San Francisco, CA, USA, 2011.
[4] D. El Baz, J. Bourgeois, Smart Cities in Europe and the ALMA Logistics Project, *ZTE Communications*, December 2015, Vol. 13, N° 4, 10 - 15.
[5] D. El Baz et al. editors, *Proceedings of the 2$^{nd}$ IEEE Smart World Congress, 13th IEEE International Conference on Ubiquitous Intelligence and Computing, 16th IEEE International Conference on Scalable Computing and Communications*, IEEE CPS, Toulouse France , July 18-21, 2016.
[6] D. El Baz, V. Boyer, J. Bourgeois, E. Dedu, K. Boutoustous, Distributed part differentiation in a smart surface, *Mechatronics*, Vol. 22, Issue 5, 2012,. 522-530.
[7] V. Boyer, D. El Baz, M. A. Salazar-Aguilar, GPU Computing Applied to Linear and Mixed Integer Programming, Chapter 10 in *Advances in GPU, Research and Practice*, H. Sarbazi-Azad editor, Morgan Kaufmann, Elsevier, Amsterdam Boston, 2017, 247 - 271.
[8] T. Fukuda, S. Nakagawa, Dynamically reconfigurable robotic system, *IEEE International Conference on. Robotics and Automation,* IEEE CPS, 1988, 1581-1586.

[9] P. M. Will, A. Castaño, W. M. Shen,. Robot modularity for self-reconfiguration, *Proceedings of the IEEE International Conference on Sensor Fusion and Decentralized Control in Robotic Systems II,* 1999, 236-246.

[10] D. Rus M. Vona, A physical implementation of the self-reconfiguring crystalline robot, *Conference on Robotics and Automation, 2000*, ICRA'00.. IEEE CPS, 2000, 2: 1726-1733.

[11] M. Yim et al. Connecting and disconnecting for chain self-reconfiguration with PolyBot[J]. *IEEE/ASME Transactions on mechatronics*, 2002, 7(4): 442-451.

[12] M. W. Jorgensen, E. H. Ostergaard, H. Lund, Modular ATRON: Modules for a self-reconfigurable robot, *Proceedings of the IEEE/RSJ International Conference on Intelligent Robots and Systems*, (IROS 2004), IEEE CPS, 2004, 2068-2073.

[13] S. C. Goldstein, J. D. Campbell, T. C. Mowry,. Programmable matter, *Computer*, 2005, 38(6): 99-101.

[14] V. Zykov et al. Robotics: Self-reproducing machines, *Nature*, 2005, 435(7039) 163-164.

[15] J. Bishop et al., Programmable parts: A demonstration of the grammatical approach to self-organization, *Proceedings of the IEEE/RSJ International Conference on Intelligent Robots and Systems 2005*. (IROS 2005), 2005: 3684-3691.

[16] H. Kurokawa, et al. Distributed self-reconfiguration of M-TRAN III modular robotic system, *The International Journal of Robotics Research*, 2008, 27 (3-4): 373-386.

[17] H. Wei et al. Sambot: A self-assembly modular robot system *IEEE/ASME Transactions on Mechatronics*, 2011, 16(4), 745-757.

[18] K. Gilpin, A. Knaian, D. Rus, Robot Pebbles: One centimeter modules for programmable matter through self-disassembly, *Proceedings of the IEEE International Conference on. Robotics and Automation* (ICRA), IEEE CPS, 2010: 2485-2492.

[19] B. T. Kirb, M. Ashley-Rollman, S. C. Goldstein, Blinky blocks: A physical ensemble programming platform, in CHI'11, *Extended Abstracts on Human Factors in Computing Systems,*.ACM, 2011, 1111-1116.

[20] D. El Baz, B. Piranda, J. Bourgeois, A distributed algorithm for a reconfigurable modular surface, *in proceedings IEEE International Parallel & Distributed Processing Symposium and Workshops* (IPDPSW 2014),. IEEE CPS, 2014, 1591-1598.

[21] J. Davey, N. Kwo, M. Yim, Emulating self-reconfigurable robots-design of the SMORES system, *IEEE/RSJ International Conference on Intelligent Robots and Systems* (IROS 2012), IEEE CPS, 2012, 4464-4469.

[22] C. Wu et al. Motion of an underwater self-reconfigurable robot with tree-like configurations, *Journal of Shanghai Jiaotong University (Science)*, 2013, 18(5): 598-605.

[23] J. W. Romanishin et al. 3D M-Blocks: Self-reconfiguring robots capable of locomotion via pivoting in three dimensions, *2015 IEEE International Conference on. Robotics and Automation* (ICRA), IEEE CPS, 2015, 1925-1932.

[24] Y. Li et al. The Kinematics Analysis of a Novel Self-Reconfigurable Modular Robot Based on Screw Theory, *DEStech Transactions on Engineering and Technology Research,* 2016.

[25] P. P. Guimarães et al. A Bio-inspired Apodal and Modular Robot, *Robotics Symposium and IV Brazilian Robotics Symposium (LARS/SBR), 2016 XIII Latin American,* IEEE CPS, 2016, 61-66.

[26] U.P. Schultz, Distributed control diffusion: towards a flexible programming paradigm for modular robots, *in: the 1st International Conference on Robot Communication and Coordination*, 2007, 15.

[27] H. Kurokawa, K. Tomita, A. Kamimura, S. Kokaji, T. Hasuo, S. Murata, Distributed self-reconfiguration of M-TRAN III modular robotic system*, Int. J. Robot. Res.* 27 (3–4), 2008 373–386.

[28] W.M. Shen, B. Salemi, P. Will, Hormone-inspired adaptive communication and distributed control for CONRO self-reconfigurable robots, *IEEE Robot. Autom. Mag.* 18 (5) (2002) 700–712.

[29] Y. Miao, G. Yan, Z. Lin, A distributed reconfiguration strategy for target enveloping with hexagonal metamorphic modules, *IEEE International Conference on Robotics and Automation* (ICRA 2011), IEEE CPS, 2011, 4804-4809.

[30] K. Boutoustous et al. Distributed control architecture for smart surfaces. 2010 IEEE/RSJ International Conference, 2010;2018-2024.

[31] DILI experiment video, https://youtu.be/kxlJRraiZQI/;2017

[32] L. De Souza Cimino, IoT and HPC integration: revision and perspectives, 2017..